\documentclass[aps,preprint,showpacs]{revtex4}
\usepackage{epsfig,amssymb,amsmath,bm}

\begin{document}

\title{
 Parity violation in deuteron photo-disintegration}

\author{M.~Fujiwara$^{ab}$ and
 A.I.~Titov$^{ac}$}
 \affiliation{
 $^a$Advanced Science Research Center, Japan Atomic Energy Research Institute,
 Tokai, Ibaraki
 319-1195, Japan\\
 $^b$Research Center of Nuclear Physics, Osaka University,
  Ibaraki, Osaka 567-0047, Japan\\
 $^c$Bogoliubov Laboratory of Theoretical Physics, JINR, Dubna 141980, Russia}

\begin{abstract}
 We analyze the energy dependence for two types of parity-non-conserving
 (PNC) asymmetries in the reaction $\gamma D\to np$
 in the near-threshold region. The first
 one is the asymmetry in reaction with circularly polarized photon beam
 and unpolarized deuteron target. The second one corresponds to those
 with an unpolarized photon beam and polarized target. We find
 that the two asymmetries have quite different energy dependence, and their
 shapes are  sensitive to the PNC-meson exchange coupling constants.
 The predictions for the future  possible experiments
 to provide definite constraints for the
 PNC-coupling constants are discussed.
\end{abstract}


 \maketitle

\section{Introduction}

 For more than forty years, the parity non-conservation
 (PNC) in nuclear processes  attracts attention as a unique tool
 for studying the strangeness conserving
 ($\Delta S=0$) weak nucleon-nucleon interaction defined
 by nontrivial interplay of the weak quark-quark interaction
 and the QCD-dynamics of composite hadrons
 at short distances~\cite{Henley69,AdelbHaxt}.
 Most of the present theoretical studies of
 the parity non-conservation in nuclear processes
 are based on the finite-range $\pi$, $\omega$
 and $\rho$-meson exchange potential
 of Desplanques, Donoghue, and Holstein (DDH)~\cite{DDH}.
 Using the symmetry consideration and the constituent quark model,
 DDH found the "reasonable range" and the "best values" of the
 the PNC meson-nucleon coupling constants.
 Their predictions are related to the theory of the weak
 interaction. Thus, the "best values" of the $\pi NN$ coupling
 for the Cabibbo
 and  Weinberg-Salam  models correspond
 to
 $h_\pi\simeq 0.2$ and 4.6 (in units of $10^{-7}$), respectively.
 The predictions for the vector meson-nucleon weak coupling
 constants are also "theory-dependent", but this dependence is not so strong.
 In case  of the charge-current theory,
 the transition  $u\to s$ responsible for the $\pi NN$ interaction
 is suppressed by  $\tan^2 \theta_C\simeq 0.05$
 ($\theta_C$ is the Cabibbo angle) as compared with the
 other transitions. This  results in strong reduction of $h_\pi$.
 The neutral-current theory is free from this
 suppression which  leads to a large value of $h_\pi$.
 The value of $h_\pi$ depends also on the non-perturbative
 QCD-dynamics of interacting mesons and baryons.
 The predictions based on
 the  Skyrmion model~\cite{Skyrmion},
 the QCD-sum rule~\cite{QCDSR},
 the soft-pion approximation~\cite{DZ},
 and the quark model with the $\Delta$ degrees of freedom~\cite{FCDH}
 give the value of $h_\pi = 0.8\sim 3\times 10^{-7}$ which is in
 the "reasonable range" of the DDH prediction (for the Weinberg-Salam model),
 being smaller than the corresponding "best value"
(see Ref.~\cite{HaxtonAn} for the review of these estimations~).

 Analysis of the available data from the nuclear PNC-experiments
 suggests that the isoscalar PNC nuclear forces dominated by the
 $\rho$ and $\omega$-meson exchange are comparable with
 the DDH "best values", whereas the isovector interaction dominated
 by the $\pi$-meson exchange is weaker by a factor of
 3~\cite{AdelbHaxt}. For example, the
 measurement of the circular polarization of the photons emitted from
 ${}^{18}$F  results in the constraint
 $0\leq h_\pi\leq 1.8 \,(\times 10^{-7})$~\cite{F18}.
 However, this constraint is in
 disagreement with the recent analysis of the ${}^{133}$Cs anapole
 moment~\cite{AnapoleCs} performed in Refs.~\cite{HaxtonAn,FlMu}.
 Quite different theoretical approaches result in similar
 conclusions: for adequate description of the data on the anapole
 moment, one needs to use
 $h_\pi$ which is about a factor of 2 greater than the DDH "best
 value" $h_\pi^{\rm best}\simeq 4.6\cdot 10^{-7}$.
 These experimental situations mentioned above impel
 the new measurement and theoretical studies to resolve
 subsisting  inconsistencies.

  The studies of the PNC-transitions in the nucleon-nucleon
  are
  very attractive because the two-nucleon wave functions are known
  reasonably well. The reactions $\gamma
  D\rightleftarrows np$ are particularly important.
  Up to now, great efforts have been devoted to
 analyzing  the thermal neutron capture by  proton in the reactions with
 unpolarized and polarized neutron.
 In this first case, the circular polarization $P_\gamma$
 of  emitted 2.23-MeV photons is analyzed.
 The experimental value
 $|P_\gamma|=(18 \pm 18)\times
 10^{-8}$~\cite{Pexp}
 is consistent with the theoretical estimations
 $|P_\gamma|=(1.8 \sim 5.6)\times
 10^{-8}$~~\cite{Danilov71,Ptheor_old,Despl80}. But, poor accuracy
 does not allow to obtain any definite conclusion about the strength of the
 PNC-forces.
 In the second case, the subject of study is the spatial
 asymmetry $A_\gamma$ of emitted photons. The experimental value of
 $A_\gamma=(6\pm21)\times 10^{-8}$~\cite{Aexp} is again too crude to
 check the theoretical predictions of $A_\gamma \sim 5\times 10^{-8}$,
( see i.g. Ref.~\cite{Desp2001}
 for reference and quotations). At present, a new
 PNC-asymmetry  measurement for the radiative neutron-proton capture is in
 preparation at LANSCE~\cite{Anew}  in order to reduce
 the experimental error of $A_\gamma$.

 Different aspects of parity non-conservation in deuteron
 electro-disintegration were analyzed in Refs.~{\cite{HH1,HH2,SCP}}.
 However, the nuclear PNC-effect in this reaction is found
 to be insignificant compared
 to the contribution of the $\gamma-Z$-boson  interference of the individual
 nucleons~\cite{SCP}.

 With the advent of the high-intensity polarized photon beams,
 investigation of PNC-effects in the $\gamma D\to np$ reaction
 becomes very important~\cite{Dreview}.
 It is clear that in this case one can
 obtain
 complementary information on the PNC-interaction. Moreover,
 one can study the dependence of the PNC-asymmetries as a function
 of  photon energy
 (contrary to the radiative np-capture, where the photon energy is fixed:
 $E_\gamma\simeq2.23$~MeV). This allows to get additional information
 which might help to reduce the ambiguity induced by uncertainties of
 the parity-conserving NN-forces at short distances. Thus, for
 example, the constraints on the PNC meson exchange coupling
 constants are usually obtained from compilation of the data extracted from
 the different experiments~\cite{HaxtonAn}. This analysis
 includes a model-dependent estimation of the PNC-matrix elements in quite
 different objects like two- and few-body systems, light
 and heavy nuclei with their own assumptions and approximations.
 The energy dependent asymmetries in the $\gamma D\to pn$ reaction
 allow to give the similar constraints using only one
 simplest nuclear system.

 In this paper, we discuss two PNC-asymmetries. One is the
 asymmetry $A_{RL}$ in  deuteron disintegration
 in the reaction with circularly  polarized photons
 and unpolarized deuteron. This asymmetry is mainly defined by the
 $\Delta I=0,2$ PNC-interaction and is equal to $P_\gamma$ at
 $E_\gamma\simeq E_{\rm thr}$, where $E_{\rm thr}$ is the
 threshold energy.
 The second one is the deuteron spin asymmetry $A_D$ in reaction with
 unpolarized beam and polarized deuteron target (polarized along-opposite
 to the beam direction). It depends also on the isovector $\Delta I=1$
 PNC-interaction, and therefore may be used for examining $h_\pi$.
 The $A_{RL}$-asymmetry was analyzed previously in
 Refs.~\cite{Lee,Oka,KK}. In Refs.~\cite{Lee,Oka}, the calculation has been
 done only with repulsive hard-core NN-potentials which
 seems to be obsolete compared to the more sophisticated
 realistic potentials with soft repulsion.
 Energy dependence of $A_{RL}$ in the region $E_\gamma-E_{\rm thr}\sim
 0.5-5$~MeV was skipped.  In Ref.~\cite{Lee}, the contribution
 of the PNC-$\pi NN$-transition was completely ignored. On the other hand,
 it was included in
 Ref.~\cite{Oka}, and the extraordinarily big contribution
 of the weak $\pi NN$ transition to $A_{RL}$ at
 $E_\gamma- E_{\rm thr}=1\sim 30$ MeV has been reported. This
 result was used by other authors (cf. e.g.~\cite{DZ,Meissner:1990ux})
 to discuss  a possibility for  extracting $h_\pi$ from
 the $A_{RL}$-asymmetry. However, in Ref.~\cite{KK}, it is shown that
 the consistent description  of all transitions defined by
  the spin-conserving $\Delta I=1$
  interaction results in their mutual cancellation which
  is a disadvantage of using $A_{RL}$ as a tool for studying
  the weak $\pi NN$ transition.
  In Ref.~\cite{KK}, the PNC-asymmetry is calculated
  on the basis of zero-range approximation where the short-range
 behaviour of the proton-neutron wave functions is modified
 phenomenologically, and therefore this result may be considered as a
 raw qualitative estimation.
 The PNC-asymmetry, $A_D$, is analyzed in Ref.~\cite{Korkin}
 within the same model as given in Ref.~\cite{KK} and therefore
 its result remains at very qualitative level.

 In our study, we use two realistic NN-potentials. One is the Paris
 potential~\cite{Paris,ParisD}
 with soft repulsion at short distances and another
 is the Hamada-Johnston (HJ) potential~\cite{HJ} with  hard core repulsion.
 The long-range meson-exchange part of the NN-interaction in these
 potentials coincides, and the difference appears at short distances.
 Our results with the  Paris potential may be useful as a prediction for
 future possible experiments, because  the Paris potential
 was designed specially for proper description of the short range
 phenomena. The results with the HJ-potential are rather
 illustrative,
 and we show them in order to link our calculation
 with the previous works and to show explicitly the effect of
 the short-range correlation as an example of the extreme  hard repulsion.

 In calculations of the PNC-asymmetries, the usage of models
 motivated by QCD (i.g., the effective chiral perturbation
 model (ChPT)~\cite{SS1,SS2}) seems to be interesting and important.
 However, the present status of ChPT  allows
 to use it only for the processes dominated by
 the long-range $\Delta I=1$ PNC forces
 (like $A_\gamma$-asymmetry~\cite{SS1}) which is not enough to be
 applied for our case where the  short-range $\Delta I=0,2$ transitions are
 important.
 Therefore, we restrict to perform the present calculation only in the
 framework of the potential description.

 This paper is organized as follows. In Section II, we define
 observables for the regular $M1$ and $E1$-transitions.
 The formula for the PNC-interactions and expressions for the
 odd-parity admixtures are given in Section III.
 In Section~IV, we discuss the results and report some predictions for the future
 experiments.  The summary is given in Section V.

\section{regular transitions}

  Near the threshold with
  $E_\gamma\leq 10$ MeV, where $E_\gamma$ is the photon energy,
  the deuteron disintegration  $\gamma D\to np$
 is dominated by the $M1$ transition $D\to{}^1S_0$ and the
  $E1$-transition $D\to {}^3P_J$.
 The amplitudes of these  $M1$ and $E1$ transitions read:
 \begin{eqnarray}
 T_{\lambda}(M1)&=& \frac{\pm i e \sqrt{k}}{2M}
 \int d{\bf r} \psi^*_f (\mu_s{\bf S} +\mu_v{\bm {\Sigma}}
 +{\bf l}_p)[{\bf n}\times
 \bm{\varepsilon}_\lambda]\psi_i,\label{M1r}\\
 T_{\lambda}(E1)&=& \frac{\pm i e \sqrt{k} }{2 }
   \int d{\bf r} \psi^*_f
  {\bf r}\bm{\varepsilon}_\lambda\psi_i,\label{E1r}
\end{eqnarray}
 where ${\bf k}={\bf n}k$ is the photon momentum,
 $\bm{\varepsilon}_\lambda$ is the photon polarization vector,
 $\lambda$ is the photon helicity,
 $M$ is the nucleon mass,
 $\mu_s=\mu_p+\mu_n=0.88$
 and $\mu_v=\mu_p-\mu_n=4.71$ are the isoscalar and isovector
 nucleon magnetic  moments, respectively; $e$ is the electric
 charge $\alpha=e^2/4\pi=1/137$;
 ${\bf r}$ is the proton-neutron relative coordinate:
 ${\bf r}={\bf r}_p-{\bf r}_n$, ${\bf l}_p$ is the proton orbital
 momentum:
 ${\bf l}_p= -i{\bf r}_p\times {\bf \nabla}_p= -i{\bf r}\times {\bf \nabla}/2
 ={\bf l}/2$;
  $\psi_i$ and $\psi_f$ are the proton-neutron wave functions
 in the initial and final states,
 defined in the obvious standard notations
 as
  \begin{eqnarray}
 \psi_{i}&=&\sum_{l \mu\sigma} \langle l\mu 1\sigma|1 M_i\rangle
 Y_{l\mu}(\hat {\bf r})\chi_{{}_{1M_i}}\frac{u_l(r)}{r},
 \nonumber\\
  \psi_{f}&=&4\pi\sum_{l s\mu\sigma} \langle l\mu s\sigma|J M_f\rangle
  Y_{l\mu}^*(\hat {\bf p}){Y}_{l\mu}(\hat {\bf r})\chi_{{}_{s\sigma}}
  \frac{u(^{2S+1}K_J:pr)}{pr},
 \label{wf}
 \end{eqnarray}
 where $u_0(r)=u(r)$ and $u_2(r)=w(r)$ are the radial
 deuteron $s$- and $d$- waves, respectively,
 and $u(^{2S+1}K_J:pr)$ ($K=S,P\ldots$) is the radial continuum wave
 function.
 The spin operators ${\bf S}$ and $\bm\Sigma$ in Eq.~(\ref{M1r})
 are defined as
 \begin{eqnarray}
 {\bf S}=       \frac12 ({\bm \sigma}_p  + {\bm \sigma}_n),\qquad
 {\bm \Sigma}=  \frac12 ({\bm \sigma}_p  - {\bm \sigma}_n).
 \end{eqnarray}
 The upper and lower signs in Eqs.~(\ref{M1r}) and (\ref{E1r})
 correspond to the photon absorption or emission, respectively~\cite{Desp2001}.

  In the following consideration, the regular and  PNC - transitions from
  the np - bound (with the radial wave function $u_D$)
  to the np ${}^3P_J$ scattering states
  (with the corresponding radial wave function $u_J$) will appear.
     Our analysis shows that these
     radial integrals at considered energies are not sensitive
     to $J$, therefore we can use "degenerated" approximation,
     where $u_J$ is calculated with the central forces.
     The reason of week sensitivity of the radial integrals to $J$
     is that the dominant contribution
     to the radial integrals comes from relatively large distances,
     where $u_{0,1,2}$ are close to each other
     because the phase shifts for different
     states at $E<10$ MeV are rather small: $|\delta_J|<4$ degrees.
     Small distances with $r<0.5$ fm, where $u_J$ are really different,
     does not contribute in the integral because $u_J$ 
     are small,
     and because of strong suppression from $u_D(r)$ (or $ru_D(r)$).
     Direct  numerical calculation shows that for $E_\gamma\lesssim 10$ MeV
     the validity of this approximation is better than 4-5\% which is
     quite reasonable. This approximation allows to express the corresponding
     matrix elements in a very transparent form useful for qualitative analysis.
     But this approximation can not be used for calculation of the odd
     parity admixtures. In this case, the spin-orbital and tensor parts
     of NN potentials have to be taken properly into account.

  The regular $M1$ and $E1$ transition amplitudes
  expressed through the radial proton-neutron wave functions have the following
  form
\begin{eqnarray}
 T_{\lambda}(M1)&=& -\lambda N\frac{\mu_v}{M} I_M^0\,\delta_{-\lambda
 M_i},\qquad I_M^0=\int u^*(^1S_0: pr)\,u(r)\,dr,
 \label{M10}\\
 T_{\lambda}(E1)&=&i N \sqrt{\frac{4\pi}{3}}
 \sum_{\mu,\sigma M_f} \langle 1\mu 1\sigma|J M_f\rangle Y^*_{1\mu}(\hat {\bf p})
 \left[\delta_{\mu\lambda}\delta_{\sigma M_i} I^0_E -\sqrt{2}
  \langle 2 m 1\sigma |1M_i\rangle\langle 2 m 1\lambda |1\mu\rangle I^2_E
  \right],\label{E10}\\
  I^0_E&=&\int u^*(^3P_J: pr)u(r)r~dr,\qquad I^2_E=\int u^*(^3P_J:
  r)w(r)r~dr,\label{E10+}
 \end{eqnarray}
 where $u(r)$ and $w(r)$ are the radial deuteron $s$- and $d$- waves,
 respectively, and $p$ is the proton momentum in c.m.s.

 The normalization factor $N$ in Eqs.~(\ref{M10}) and (\ref{E10})  reads
 \begin{eqnarray}
 N^2=\frac{2\alpha\pi k}{p^2}.
 \end{eqnarray}

 The total cross section is related to the amplitudes $T_\lambda$
 as
 \begin{eqnarray}
 \sigma^{\gamma D\to np}&=& \frac{M p}{12\pi}\sum_{\lambda M_i}
 \left(\overline{|T_\lambda(M1)|^2} \,
 + \, \overline{|T_\lambda(E1)|^2}\right),
 \label{sigma_tot}\\
 \qquad \overline{|T_\lambda|^2}&=&\frac{1}{4\pi}\int d\Omega_p
  {|T_\lambda|^2},
 \label{T**2av}
 \end{eqnarray}
where  $M_i$ is the deuteron spin projection and
 \begin{eqnarray}
 \frac{1}{2N^2} \sum_{\lambda M_i } \overline{|T_\lambda(M1)|^2} &=&
 \left(\frac{\mu_v}{M}\right)^2 \vert I_M^0\vert^2,\nonumber\\
 \frac{1}{2N^2} \sum_{\lambda M_i} \overline{|T_\lambda(E1)|^2} &=&
 \vert I^0_E\vert^2 + \frac{2}{5} \vert I^2_E\vert^2.
 \label{T**2}
 \end{eqnarray}
In the following, we will assume the average of Eq.~(\ref{T**2av})
in all quadratic forms of $T_aT_b^*$ which define the observables
for the case when the angular distribution of the final nucleon is
not fixed and skip the symbol "overline",  for simplicity.

 The wave functions for
 the deuteron bound state and the $np$-scattering states
 are calculated using the
 realistic nucleon-nucleon potentials for two extreme
 cases:
  potential with soft short-range repulsive core
 (Paris potential~\cite{Paris,ParisD}) and potential with hard-core repulsion
 (Hamada-Johnston (HJ) potential~\cite{HJ}).

 Figure~\ref{fig:2} shows the result of our calculation for the
 total cross section of the $\gamma D\to np$ reaction as a function
 of the energy excess:
 $\Delta E_\gamma=E_\gamma-E_{\rm thr}$, where $E_{\rm thr}$ is
 the threshold energy
 $E_{\rm thr}=\epsilon(1+\epsilon/2(M_p+M_n-\epsilon))\simeq\epsilon$, and
 $\epsilon=2.23$ MeV is the deuteron binding energy, together with
 available data~\cite{Barnes,Birenbaum}.
 The result for the Paris potential is shown in
 Fig.~\ref{fig:2}~(a), where
 each  contribution from $M1$ and $E1$ transitions is also displayed.
 The difference
 between the Paris and HJ-potentials in the total cross section
 does not exceed 5\% and disappears at $\Delta E_\gamma\to 0$
 (cf. Fig.~\ref{fig:2}~(b)) because the
 main contribution into the radial integrals of
 Eqs.~(\ref{M1r}) and (\ref{E1r}) at small $\Delta E_\gamma$
 comes from the relatively large distances
 with $r\gg1$ fm, where the $np$-wave functions calculated
 for all the realistic potentials are close to each other.
 This result is in agreement with those of the previous calculations performed
 with various realistic potentials (cf. Ref.~\cite{Birenbaum} for references
 and quotations).

\section{PNC-interaction and parity odd admixtures}

The short range PNC  potential is expressed in terms of
$\rho,\omega$ and $\pi$ exchanges and has the following form
~\cite{DDH,Henley78}
\begin{eqnarray}
 V_{\rm PNC}= &&\frac{2ig_\rho}{M}
 \left\{\left[h^0_\rho {\bm \tau}_1 {\bm \tau}_2
  +\frac12h^1_\rho (\tau_1^z+ \tau_2^z)
  +\frac{1}{2\sqrt{6}} h^2_\rho(3\tau_1^z\tau^z_2-{\bm \tau}_1 {\bm
  \tau}_2)\right]\right.  \nonumber\\
 &&~~~~~~ \times  \left( {\bm \Sigma} \{ {\bf \nabla},f_\rho(r)\}
  +(1+\chi_\rho)\,{\bm \Omega} {\bf \nabla}
    f_\rho(r)\right)\nonumber\\
 &&~~~~~~ - \left. \frac12h^1_\rho (\tau_1^z - \tau_2^z)\,\,
   {\bf S}\, \{ {\bf \nabla}, f_\rho(r)\}
  +i h^{1'}_\rho \left[\frac{{\bm\tau_1}\times{\bm\tau_2}}{2}\right]^z\,\,
   {\bf S}\,{\bf \nabla}\,f_\rho(r) \right\}
  \nonumber\\ \nonumber\\
 &+& \frac{2ig_\omega}{M}\left\{
 \left[ h^0_\omega +\frac12h^1_\omega (\tau_1^z+
 \tau_2^z)\right]
 \left( {\bm \Sigma} \{ {\bf \nabla},f_\omega(r) \}
  +(1+\chi_\omega)\, {\bm \Omega} {\bf \nabla}
  f_\omega(r)\right)\right.
  \nonumber\\
 &&~~~~~~+\left. \frac12h^1_\omega (\tau_1^z - \tau_2^z)\,\,
   {\bf S}\,\{ {\bf \nabla}, f_\omega(r)\} \right\}\nonumber\\
  \nonumber\\
 &+&\frac{2g_\pi h_\pi}{\sqrt{2}M}
 \left\{\left[ \frac{{\bm\tau}_1\times {\bm\tau_2}}{2}\right]^z\,
  {\bf S\, \nabla} h_\pi(r)\right\},
 \label{V-pnc}
 \end{eqnarray}
where
 \begin{eqnarray}
 f_{\omega}(r)\simeq f_\rho(r)=\frac{{\rm e}^{-{m_\rho r}}}{4\pi r},\qquad
 h_\pi(r)=\frac{{\rm e}^{-{m_\pi r}}}{4\pi r}, \qquad{\bm \Omega}=
 \frac{i}{2}[\bm \sigma_1\times \bm \sigma_2].
 \end{eqnarray}
For the strong nucleon-meson coupling constants $g_i$ and
$\chi_i$, we use commonly accepted values~\cite{Adelb83}:
$g_\rho=2.79, g_\omega=8.37, g_\pi=13.45$, $\chi_\rho=3.71$
$\chi_\omega=-0.12$. The PNC meson-nucleon coupling constants,
$h_i$, are taken as the "best value" of Ref.~\cite{DDH} for the
Weinberg-Salam model.  The sensitivity of the observables to
$h_\pi$ will be discussed separately.  For convenience, Table~1
shows all parameters used in the present work.
Parity-odd admixture states $\widetilde\psi$ to the deuteron wave
functions and np-scattering states are defined in the first order
of perturbation theory in terms of Schr\"odinger equation
 \begin{eqnarray}
  [E-H_{\rm PC}]\widetilde\psi=V_{\rm PNC}\psi,
   \label{SE}
 \end{eqnarray}
 where $H_{\rm PC}$ is the parity-conserving Hamiltonian and $V_{\rm PNC}$
 is the parity-violating two-body potential. For the odd-parity
 $^1P_1$ admixture in a deuteron with $I=0$, we have the following expression:
 \begin{eqnarray}
 \widetilde\psi(^1P_1) &=& i\frac{\widetilde u(^1P_1:r)}{r}
 Y_{1M_i}(\hat {\bf r }){\chi_{}}_{00},
 \nonumber\\
 \widetilde u(^1P_1:r)&=&\sum_{i=\omega,\rho} \frac{2g_i\hat h^0_i}{\sqrt{3}}
 \int dr'\,g^{00}_1(-\epsilon;r,r')\left\{\left[-\chi_i f'_i(r')
  +2f_i(r')(\frac{\partial}{\partial
  r'}-\frac{1}{r'})\right]u(r')\right.\nonumber\\
  &&\qquad-\left.\sqrt{2}\left[-\chi_i f'_i(r')
  +2f_i(r')(\frac{\partial}{\partial r'}
  +\frac{2}{r'})\right]w(r')\right\},\nonumber\\
  \hat h^0_\rho&=& - 3 h^0_\rho,\qquad \hat h^0_\omega= h^0_\omega,
  \label{1P1}
 \end{eqnarray}
 where ${\chi_{}}_{SS_z}$ is the two nucleon spin function,
 $g_l^{IS}(E;r,r')$ is the Green function of
 the radial Sch\"odinger equation for the $np$-system with the orbital
 momentum $l=1$, isospin $I=0$, spin $S=0$ and the energy
 $E=-\epsilon$.

 The odd-parity $^3P_1$ admixture with $I=1$ is dominated by the $\pi$-meson
 exchange weak interaction. Nevertheless, for completeness
 we also include
 the contributions of  the $\rho$
 and $\omega$-meson exchanges for $\Delta I=1$ transition.
 The net expression for the $^3P_1$ admixture reads
 \begin{eqnarray}
 \widetilde\psi(^3P_1) &=&i\sum_{\mu\sigma}\langle 1\mu 1\sigma|1M_i\rangle
 \frac{\widetilde u(^3P_1:r)}{r}Y_{1\mu}(\hat {\bf r}){\chi_{}}_{1\sigma},
 \nonumber\\
 \widetilde u(^3P_1:r)&=& \frac{2}{\sqrt{3}}
 \int dr'\,g^{01}_1(-\epsilon;r,r')\left\{
 \left(g_\pi h_\pi\, f'_\pi(r')- \sqrt{2} g_\rho {h'}^{1}_\rho
 f'_{\rho}(r')\right)
 \left[u(r')+\frac{1}{\sqrt{2}}w(r')\right]\right.\nonumber\\
 &-&\sqrt{2} (g_\omega h^1_\omega - g_\rho h^1_\rho )
 \left[(f'_{\rho}(r') + 2f_{\rho}(r')(\frac{\partial}{\partial r'}-\frac{1}{r'}))
 u(r')\right.\nonumber\\
  &&\qquad\qquad \left. \left.
  + \frac{1}{\sqrt{2}}(f'_{\rho}(r') + 2f_{\rho}(r')(\frac{\partial}{\partial r'}
  +\frac{2}{r'}))w(r') \right]\right\}.
 \label{3P1}
 \end{eqnarray}

 Figure~\ref{fig:3}~(a)
 shows the odd-parity ${}^1P_1$ and ${}^3P_1$ admixture in the
 deuteron wave function
 for the Paris (solid curves) and HJ (dashed curves) potentials.
 The main difference between the two potentials appears
 at short distances. In case of the HJ potential,
 all wave functions vanish in
 the core-region with $r\leq r_{\rm core}$ ($r_{\rm core}=0.48$ fm).
 This results in a sizeable suppression
 of  ${}^1P_1$-admixture because the "form factors" $f_v(r)$
 in Eq.~(\ref{1P1}) decrease sharply with $r$. The function
 $h_\pi(r)$ decreases more slowly. Therefore, the ${}^3P_1$-admixture
 is not so sensitive to the choice of the potential model.

 Analysis of the odd parity component in the continuum np-states
 shows that at $E_\gamma<10$ MeV, the dominant contribution to
 the considered asymmetries comes
 from the $^3P_0$ admixture to the $^1S_0$ state, from the  $^1S_0$ admixture to the
 $^3P_0$ state, and from
 the ${}^3S_1$ and  ${}^3D_1$-components of the  $^3P_1$-state.
 They are defined as follows
 \begin{eqnarray}
 \widetilde\psi(^3P_0) &=& i\frac{\sqrt{{4\pi}}}{3}
 \sum_{\mu}
 \frac{\widetilde u(^3P_0:pr)}{pr}
 (-1)^{\mu+1}Y_{1\mu}(\hat {\bf r}){\chi_{}}_{1-\mu},
 \label{3P0}\\
 \widetilde u(^3P_0:pr)&=-&\sum_{i=\rho,\omega}{\sqrt{12} g_i \widehat h_i}
 \int dr'\,g^{11}_1(E;r,r')\nonumber\\
 &\times& \left[(2+\chi_i) f'_i(r')
  + 2f_i(r')(\frac{\partial}{\partial
  r'} -  \frac{1}{r'})\right]u(^1S_0:pr'),
 \nonumber\\
 \nonumber\\
 \widetilde\psi(^1S_0) &=& i\sqrt{\frac{4\pi}{3}}
 \frac{\widetilde u(^1S_0:pr)}{pr}{\chi_{}}_{00}
 \sum_m Y_{1m}^*(\hat {\bf p}),
  \label{1S0}\\
 \widetilde u(^1S_0:pr)&=&\sum_{i=\rho,\omega} \frac{2g_i\widehat h_i}{\sqrt{3}}
 \int dr'\,g^{10}_0(E;r,r')\left[\chi_i f'_v(r')
  -2f_i(r')(\frac{\partial}{\partial
  r'}+\frac{1}{r'})\right]u(^3P_0:pr'),
  \nonumber\\
\widetilde\psi(^3S_1) &=& i\sqrt{{4\pi}}\frac{\widetilde
 u(^3S_1:pr)}{pr}{\chi_{}}_{1M_f}
 \sum_m Y_{1m}^*(\hat {\bf p}),
 \label{3S1}\\
 \widetilde u(^3S_1:pr)&=&-\frac{2}{\sqrt{3}}
 \int dr'\,g^{01}_0(E;r,r')\left[
  g_\pi h_\pi\, f'_\pi(r')- \sqrt{2} g_\rho {h'}^{1}_\rho
 f'_{\rho}(r') \right. \nonumber\\
 &+&\sqrt{2} (g_\omega h^1_\omega - g_\rho h^1_\rho )
 \left.\left( f'_{\rho}(r') + 2f_{\rho}(r')(\frac{\partial}{\partial r'}
 +\frac{1}{r'})\right){}\right] u(^3P_1:pr'),
\nonumber\\
 \nonumber\\
 \widetilde\psi(^3D_1) &=& i\,{4\pi}\frac{\widetilde u(^3D_1:pr)}{pr}
 \sum_{\mu\sigma}\langle 2\mu 1\sigma|1M_f\rangle
  Y_{2\mu}(r){\chi_{}}_{1\sigma}
  \sum_m Y^*_{1m}(\hat {\bf p}),
  \label{3D1}\\
 \widetilde u(^3D_1:pr)&=&-\sqrt{\frac{2}{3}}
 \int dr'\,g^{01}_2(E;r,r') \left[
 g_\pi h_\pi\, f'_\pi(r')- \sqrt{2} g_\rho {h'}^{1}_\rho
 f'_{\rho}(r') \right. \nonumber\\
 &+&\sqrt{2} (g_\omega h^1_\omega - g_\rho h^1_\rho )
 \left.\left( f'_{\rho}(r') + 2f_{\rho}(r')(\frac{\partial}{\partial r'}
 -  \frac{2}{r'})\right)\right] u(^3P_1:pr'),
 \nonumber\\
  \widehat h_\rho&=&h^0_\rho-\sqrt{\frac{2}{3}} h^2_\rho,
  \qquad \widehat h_\omega= h^0_\omega,
  \nonumber
 \end{eqnarray}
 where $E=p^2/M$.
 The Green functions $g(E;r,r')$ in Eqs.(\ref{1P1})-(\ref{1S0}) are
 expressed through the regular and irregular solutions of the
 corresponding Schr\"odinger equations in the standard way.
 For the $^3S_1$ and  $^3D_1$ states we use their
 spectral representation
 \begin{eqnarray}
 Mg^{01}_l(E;r,r')=\frac{u_l(r)u_l(r')}{E+\epsilon}
 +\frac{2}{\pi}\int  d\,k\,
 \frac{u(^3K_1:kr)u(^3K_1:kr')}{E-E_{\bf k}},
 \label{g11}
 \end{eqnarray}
 with $\int u_l^2\,dr=1$,  $K=S,D$ and $E_k=k^2/M$,  and keeping only the first term,
 because the second term does not
 contribute to the $M1$-transition. In this sense, our $^3S_1$, $^3D_1$-odd
 parity admixtures are only the part of the corresponding
 total wave functions which contribute to the PNC $M1$-transition.

  Figure~\ref{fig:3}~(b) shows
   the odd parity ${}^3P_0$, ${}^1S_0$, ${}^3S_1$ and  ${}^3D_1$-
  admixtures for two potentials at $\Delta E_\gamma=0.1$ MeV.
  The ${}^3D_1$-function is
  scaled additionally by $\sqrt{P_D}$, where $P_D$ is the
  probability of the $D$-state in a deuteron, because
  the corresponding $M1$-transition is suppressed by this
  factor ($P_D^{\rm Paris}=0.0577$, $P_D^{\rm HJ}=0.0697)$.
  Again, one can see that in
  case of hard-core potentials, all wave functions vanish in
  the core-region, which leads to the relative suppression of the
  odd-parity ${}^3P_0$ and ${}^1S_0$-components, whereas
  the ${}^3S_1$ and  ${}^3D_1$-configurations defined mainly by
  the long-range $\pi NN$ interaction
  are not sensitive to the potential at $r>r_{\rm core}$.
  In Fig.~\ref{fig:3}~(c), we show the  continuum wave functions at
  $\Delta E_\gamma=1$ MeV. The main difference as compared with the previous
  case appears in the $^1S_0$ odd-parity admixture.
  It oscillates with $r$ more strongly and
  has a node at $r\simeq3.5$ fm at  $\Delta E_\gamma=1$ MeV. This
  oscillating behaviour is manifested in the corresponding $M1$-transition.

\section{Asymmetries}

The asymmetry of the deuteron disintegration in reaction with
circularly polarized photon beam
\begin{equation}\label{Alr}
 A_{RL}=\frac{\sigma_{\lambda=1} - \sigma_{\lambda=-1}}
             {\sigma_{\lambda=1} + \sigma_{\lambda=-1}},
\end{equation}
consists of seven terms
\begin{eqnarray}
 A_{RL}=
 \sum_{i=1}^4  {V}_i^\gamma +
       \sum_{j=1}^3 {\pi}_j^\gamma,
 \label{Alr2,}
\end{eqnarray}
 defined by the interplay of dipole transitions caused by the
 parity-conserved and parity non-conserved interaction
 as follows
  \begin{subequations}
 \label{Alr2}
 \begin{eqnarray}
 {V}_1^\gamma   &=&2\,{\rm Re}\left[T^*(M1:D\to {^1S_0} )
 \,{T}(E1:D\to\widetilde{^3P_0})
 \right] /{\cal N},  \label{Alr2-1}\\
 {V}_2^\gamma    &=& 2\,{\rm Re}\left[ T^*(M1:D\to{^1S_0})
 \,{T}(E1:\widetilde{^1P_1}\to{}^1S_0)
 \right] /{\cal N},\label{Alr2-2}\\
 {V}_3^\gamma   &=&2\,{\rm Re}\left[ T^*(E1:D\to {^3P_0})
 \,{T}(M1:D\to\widetilde{^1S_0})
 \right]/{\cal N},\label{Alr2-3}\\
 {V}_4^\gamma    &=& 2\,{\rm Re}\left[ T^*(E1:D\to {^3P_J})
 \,{T}(M1:\widetilde{^1P_1} \to{}^3P_J)
 \right] /{\cal N},\label{Alr2-4}\\
 {\pi}_1^\gamma &=&2\,{\rm Re}\left[ T^*(E1:D\to {^3P_J})
 \,{T}(M1:\widetilde{^3P_1} \to{}^3P_J)
 \right]/{\cal N},\label{Alr2-5}\\
 {\pi}_2^\gamma  &=& 2\,{\rm Re}\left[ T^*(E1:D\to {^3P_1})
 \,{T}(M1:D\to\widetilde{^3S_1})
 \right]/{\cal N},
  \label{Alr2-6}\\
 {\pi}_3^\gamma  &=& 2\,{\rm Re}\left[ T^*(E1:D\to {^3P_1})
 \,{T}(M1:D\to\widetilde{^3D_1})
 \right]/{\cal N},
  \label{Alr2-7}\\
 {\cal N}&=&\frac{1}{2N^2} {\rm Tr}\,[TT^*].
  \nonumber
 \end{eqnarray}
  \end{subequations}
Their explicit form in terms of the radial integrals read
 \begin{subequations}
 \label{Alr3}
 \begin{eqnarray}
 {V}_1^\gamma   &=&-\frac{2}{3\sqrt{3}}\frac{1}{{\cal N}}\frac{\mu_v}{M}
 \,{\rm Re}\left[ {I_M^0}^*\cdot
 \int dr \,r \widetilde{u}^*(^3P_0:pr)\,[u(r)-\sqrt{2}w(r)]
 \right],
 \label{Alr3-1}\\
 {V}_2^\gamma   &=&-\frac{2}{\sqrt{3}}\frac{1}{{\cal N}}\frac{\mu_v}{M}
 \,{\rm Re}\left[{I_M^0}^*
 \cdot \int dr \,r {u}^*(^1S_0:pr)\,\widetilde{u}({}^1P_1:r)
 \right], \label{Alr3-2}\\
 {V}_3^\gamma   &=& \frac{2}{3\sqrt{3}}\frac{1}{{\cal N}}\frac{\mu_v}{M}
 \,{\rm Re}\left[ \left({I_E^0}^* -\sqrt{2}{I^2_E}^* \right)
 \cdot \int dr \, \widetilde{u}^*(^1S_0:pr)\,u(r)
 \right], \label{Alr3-3}\\
 {V}_4^\gamma   &=& \frac{2}{\sqrt{3}}\frac{1}{{\cal N}}\frac{\mu_v}{M}
 \,{\rm Re}\left[\left( {I_E^0}^* - \sqrt{2}{I^2_E}^*\right)
 \cdot \int dr \, {u}^*(^3P_J:pr)\,\widetilde{u}(^1P_1:r)
 \right], \label{Alr3-4}\\
 {\pi}_1^\gamma   &=& - \sqrt{\frac83}\frac{1}{{\cal N}}
 \frac{\mu_s}{M}
 \,{\rm Re}\left[\left({I_E^0}^*+\frac{1}{\sqrt{2}}{I^2_E}^* \right)
 \cdot \int dr \, {u}^*(^3P_J:pr)\,\widetilde{u}(^3P_1:r)
 \right], \label{Alr3-5}\\
 {\pi}_2^\gamma   &=& \sqrt{\frac83}\frac{1}{{\cal N}}\frac{\mu_s}{M}
 \,{\rm Re}\left[ \left({I_E^0}^*+\frac{1}{\sqrt{2}}{I^2_E}^*\right)
 \cdot \int dr \, \widetilde{u}^*(^3S_1:pr)\,u(r)\right],
  \label{Alr3-6}\\
 {\pi}_3^\gamma   &=& -\sqrt{\frac{2}{3}}\frac{1}{{\cal N}}\frac{\mu_s-3/2}{M}
 \,{\rm Re}\left[\left({I_E^0}^*+\frac{1}{\sqrt{2}}{I^2_E}^*\right)
 \cdot \int dr \, \widetilde{u}^*(^3D_1:pr)\,w(r)
 \right]. \label{Alr3-7}
\end{eqnarray}
  \end{subequations}
 Another asymmetry is related to the deuteron
disintegration with unpolarized photon beam and polarized deuteron
target
\begin{equation}\label{AD}
 A_{D}=\frac{\sigma_{M_D=1}  - \sigma_{M_D=-1}}
            {\sigma_{M_D=1}  + \sigma_{M_D=-1}},
\end{equation}
where $M_D=1(-1)$ corresponds to the deuteron spin projection
parallel (antiparallel) to the direction of the beam momentum.
This asymmetry has also seven components
\begin{eqnarray}
 A_{D}=
 \sum_{i=1}^4  {V}_i^D +
       \sum_{j=1}^3 \,{\pi}_j^D.
 \label{AD2,}
\end{eqnarray}
Three of them,  $V^D_{1,2,3}$, are equal with the opposite sign to
the corresponding $V^\gamma\,-$ asymmetries
 \begin{eqnarray}
 {V}_1^D   &=&-{V}_1^\gamma,\qquad  {V}_2^D=-{V}_2^\gamma,\qquad
 {V}_3^D   = -{V}_3^\gamma.
 \label{AD1-3}
\end{eqnarray}
 In these cases, the spin of the final states is zero
 and the corresponding $M1$-transitions are proportional
 to $\delta_{-\lambda M_D}$.
 The other four asymmetries are expressed as
 \begin{eqnarray}
 {V}_4^D   &=& \frac{2}{\sqrt{3}} \frac{1}{{\cal N}}\frac{\mu_v}{M}
 {\rm Re}\left[ \left({I_E^0}^* - \sqrt{2}{I^2_E}^* \right)
 \cdot \int dr \, {u}^*(^3P_J:pr)\,\widetilde{u}(^1P_1:r)\right], \nonumber\\
 {\pi}_1^D   &=& -\sqrt{\frac23}\frac{1}{{\cal N}}
 {\rm Re}\left[\left(\frac{\mu_s-1}{M} {I_E^0}^*
 -\sqrt{2}\frac{\mu_s-1/4}{M}{I^2_E}^*\right)\cdot
 \int dr \, {u}^*(^3P_J:pr)\,\widetilde{u}(^3P_1:r)\right], \nonumber\\
 {\pi}_2^D   &=& -\frac{1}{2} {\pi}_2^\gamma,
 \qquad{\pi}_3^D   = -\frac12 {\pi}_3^\gamma.
 \label{AD4-6}
\end{eqnarray}
The most important is the modification of ${\pi}_1^D$. As we will
see later, the spin-transitions in ${\pi}_1$ and ${\pi}_2$
proportional to $\mu_s$ are almost canceled in $A_\gamma$, but not
in $A_D$. Therefore, the PNC-weak interaction of the
$\pi$-exchange may be clearly manifested only in the
$A_D$-asymmetry.

\section{Results and discussion}

 We first discuss the $A_\gamma$-asymmetry.
 At  $E_\gamma\to E_{\rm thr}$,  the $V_1$- and $V_2$- terms only  contribute to
 the total asymmetry. The signs of them are opposite and therefore
 their  interference is destructive. The sign of the total
 asymmetry is defined by the dominant term. The strength of
 $V_{1,2}$ is determined by the values of the corresponding PNC-weak
 coupling constants and the behaviour of the proton-nucleon wave
 functions at short distances. For the case when the functions
 $u(r)$ and
 $u(^1S_0:pr)$ are smooth  at $r\lesssim 1$ fm (e.g. in the zero range
 approximation), one can neglect derivatives $u'$ in
 Eqs.~(\ref{1P1}) and (\ref{3P0}). Using the approximate
 expression for the Green function for $r'<r$ and $E\sim 0$: $g_1(E: r,r')
 \simeq
  -{r'}^2\theta(r-r')/3r$, neglecting $w$ and $w'$,
 and taking into account the fact that
 the main contribution to the odd parity admixtures
 $\widetilde{u}(^1P_1:r)$  and
 $\widetilde{u}(^3P_0:pr)$ comes from the
 terms proportional to $f'_{v}(r')$, one gets the following
 estimate
 \begin{eqnarray}
 \frac{V_1^\gamma}{V_2^\gamma}
 \simeq - \frac{(h^0_\rho-\sqrt{\frac{2}{3}}h^2_\rho)(2+\chi_\rho)+
 h^0_\omega(2+\chi_\omega)}
 {3 h^0_\rho\chi_\rho - h^0_\omega\chi_\omega}\simeq-0.18.
 \label{V1V2}
 \end{eqnarray}
 This estimation coincides with the result of the plane-wave Born
 approximation given in Ref.~\cite{AdelbHaxt} and shows the dominance of
 the $^3S_1\to{}^1\widetilde{P_1}$ PNC-transition with  $\Delta I=0$
 compared to the $^1S_0\to {}^3\widetilde{P_0}$
 with $\Delta I=0,2$. In case of the realistic
 NN-potential, the radial $np$-wave functions increase
 rapidly from zero at $r=0$
 (for the hard core potential from $r=r_{\rm core}$)
 to the finite value at $r\simeq 1$~fm.
 Since $f_{v}$ and $|f'_{v}|$ decrease with $r$,
 the dominant contribution
 to the integrals in Eqs.~(\ref{1P1}) and (\ref{3P0})
 comes from the regions of $r = 0.6\sim 1.2$~fm.  This leads
 to increase of
 $|{V_1^\gamma}/{V_2^\gamma}|$ and to decrease of the
 asymmetries
 $|A_{RL}|$ and $|A_D|$. Of course, we can not neglect the terms
 with derivatives $u'$ because they
 are essential just in
 the region of the dominant contribution of the corresponding integrals.
 In our case $u'(r),\,w'(r),\, u'({}^1S_0:r)$
 at $r\lesssim 1.2$ fm
 are positive and large, especially
 for the hard-core (HJ)-potentials.
 In Eq.~(\ref{1P1}),
 the term proportional to $u'(r)$
 gives a constructive contribution and enhance $|V_2|$,
 whereas in Eq.~(\ref{3P0}),
 $u'_{np}(pr)$ contributes destructively and suppresses $|V_1|$. As
 a result, we get the ratio of  ${V_1^\gamma}/{V_2^\gamma}$ close to its
 raw estimate of Eq.~(\ref{V1V2}).

 Figure~\ref{fig:7}(a) shows the asymmetries $A_{RL}$
 as a function of $\Delta E_\gamma$ together with
 the partial asymmetries $V_i$ and $\pi_i$.
 When $\Delta E_\gamma$ increases, the PNC $M1$ transitions become
 important. At low $\Delta E_\gamma$,  asymmetries $V^\gamma_3$
 caused by the
  $\Delta I=0,2$ PNC-forces  and  $V^\gamma_4$, generated by $\Delta
 I=0$ forces  are close to each other numerically
  with the same sign. However, at
 $\Delta E_\gamma\sim 0.5$~MeV,  $V^\gamma_3$ decreases, changes sign
 and then its absolute value becomes much smaller than  $|V^\gamma_4|$, and
 it does not affect the asymmetry. In the limit of $\Delta E_\gamma\to 0$
 our result ($A_{LR}=3.35\times 10^{-8}$) is in agreement with the
 previous calculations of the circular photon polarization in
 the $np\to D\gamma$ reaction ($P_\gamma=(1.8 \sim 5.6)\times
 10^{-8}$~~\cite{Danilov71,Ptheor_old,Despl80}.

 The PNC transitions with $\Delta I=1$ ($\Delta S=0)$ are described by the
 $\pi_1^\gamma$, $\pi_2^\gamma$ and $\pi^\gamma_3$-terms, where
 $\pi^\gamma_{1,2}$ terms are dominant and they are mostly determined by the weak
 $\pi$-meson exchange interaction. In
 Fig.~\ref{fig:7}(a),  we show the $\pi^\gamma_1$-asymmetry, the sum of $\pi_2^\gamma +
 \pi^\gamma_3$-terms,
 and the coherent sum of all the  $\Delta I=1$ transitions denoted as
 $\pi^\gamma$.  At $\Delta E\sim 10$ MeV,  the absolute values of
 $\pi^\gamma_{1}$  and $\pi^\gamma_{2}$ are the biggest among the other ($V_i$) terms
 and close to each other. But their signs are opposite. Therefore, the coherent sum
 is rather small
 \begin{eqnarray}
\pi^\gamma_{12} = \pi^\gamma_{1} + \pi^\gamma_{2}\sim
\mu_s(\widetilde {I_M^1} - \widetilde {I_M^2})\sim
\mu_s\textit{O}(P_D),
\end{eqnarray}
where $\widetilde{I_M^1}$ and $\widetilde{I_M^2}$ are the radial
integrals for the $M1$-transitions in Eqs.~(\ref{Alr3-5}) and
(\ref{Alr3-6}), respectively.
 The finite value of $\pi^\gamma_{12}$ is mainly caused by the non-symmetrical
contribution of the deuteron $d$-wave in $\pi^\gamma_1$,  and
$\pi^\gamma_2$ and it almost vanishes when $P_D=0$. In case of the
zero range approximation in the limit $\Delta E_\gamma\to 0$, this
cancellation is exact~\cite{KK}. In the real case the total
contribution of the $\Delta I=1$ PNC interaction ($\pi^\gamma$) is
finite. However, its absolute value is smaller by a factor of 27
as compared with the result of Ref.~\cite{Oka}. Therefore, it
seems to be difficult to get information about the $\Delta I=1$
PNC forces from $A^\gamma_{RL}$

 The coherent interference of the $V_1^\gamma$- $V_2^\gamma$- and
 $V_4^\gamma$-terms leads to  sharp decrease of $A_{RL}$ down to zero at
 $\Delta E_\gamma\simeq 1.3$~MeV (in case of Paris potential),
 and change a sign from positive to negative.
 Figure~\ref{fig:7}~(b) shows the total asymmetry $A_{RL}$ for the two
  potentials. For illustration,  we also show
  the prediction of Ref.~\cite{KK}
  for the modified ZRA-model. One can see that
  the behaviour of the asymmetry $A_{RL}$ is similar qualitatively for the
  quite different models.
  In case of the HJ-potential, the asymmetry is smaller.
  The difference between two potentials
  at small $\Delta E_\gamma=0.01\sim 1$ MeV
  amounts to a factor of $2.5\sim 3$.
  The intercept $A_{RL}=0$  is shifted towards
  lower energies.
  The prediction of the modified ZRA-model~\cite{KK} is close qualitatively
  to those of the Paris potential but the absolute value of $A_{RL}$
  is much greater and the position of the intercept is shifted towards higher
  energies. This comparison with HJ-potential and ZRA-model
  has a rather illustrative character
  because the realistic potentials  with the soft core repulsion
  are commonly accepted to be more adequate
  for description of the short range phenomena. From this point of
  view,  only the prediction obtained with Paris potential seems to be realistic.

 Figure~\ref{fig:8}(a)
 shows the $A_D$-asymmetry as a function of $\Delta
 E_\gamma$. There are two main differences compared to the $A_{RL}$
 asymmetry. Firstly,  the components $V_2$ and $V_4$ are of the same sign.
 Secondly, there is no cancellation  between the $\widetilde{^3P}_1\to{}^3P_J$ and
 $D\to \widetilde{{}^3S}_1$ - transitions. Their coherent sum now
 behaves  as
\begin{eqnarray}
\pi^D_{12}= \pi^D_{1} + \pi^D_{2}\sim
(\mu_s-\frac12)\widetilde{I_M^1},
\end{eqnarray}
 and becomes a significant part of the asymmetry  at large $\Delta
 E_\gamma$.
 The sum of all transitions  generated by  the $\Delta I=1$ PNC
 forces $\pi^D=\pi^D_{12}+\pi_3^D$ has the same sign as the
 $V_2$- and $V_4$-components. This leads to a non-monotonical behaviour
 of $|A_{D}|$ with a local minimum at $\Delta E_\gamma\simeq 2$~MeV, but
 the sign of $A_D$ remains to be the same at $0<\Delta E_\gamma\leq 10$ MeV
 and negative.
 In Fig.~\ref{fig:8}(b), we compare the results for $A_D$
 calculated with the two potentials.
 The difference between two asymmetries decreases with increasing the photon energy.
 However, the two results are similar in shape.

 The weak $\pi$-meson exchange is mostly important
 at large $\Delta E_\gamma$.
 For illustration, Fig.~\ref{fig:9} shows the asymmetry $A_D$
 calculated as a function of $\Delta E_\gamma$ at different values
 of $h_\pi$ which cover its theoretical uncertainty:
 $0\leq h_\pi \leq 2.5 h_\pi^{\rm best}$,
 where $h_\pi^{\rm best}$ is the "best" value of DDH.
 One can see that the constructive interference between
 week $\pi$, and vector meson exchange
 results in increasing
 the absolute value of $A_D$ with increasing  $h_\pi$ and leads to
 shift the position of the local minimum towards the lower energies.
 The absolute
 value of $|A_D|$ increases by a factor of $3$
 when $R_\pi$ changes from 0 to 2.5
 at $1 \lesssim\Delta E_\gamma \lesssim 10$~MeV.

 Using the energy dependence of $A_{RL}$ and $A_{D}$,
 one can obtain relations between
 the weak coupling constants.
 Thus, the standard representation of asymmetries through
 $h_i$ and $h_\pi$ read
 \begin{eqnarray}
 A_{RL}&=&a_\rho^0 g_\rho h_\rho^0
 +a_\rho^2 g_\rho h_\rho^2
 +a_\omega^0 g_\omega h_\omega^0
 +a^1_v(g_\omega h_\omega^1 -g_\rho h_\rho^1 )
 +{a'}^1_\rho g_\rho {h'_\rho}^1
 +a_\pi g_\pi h_\pi,
 \label{ALR-s}\\
 A_{D}&=&b_\rho^0 g_\rho h_\rho^0
 +b_\rho^2 g_\rho h_\rho^2
 +b_\omega^0 g_\omega h_\omega^0
 +b^1_v(g_\omega h_\omega^1 -g_\rho h_\rho^1 )
 +{b'}^1_\rho g_\rho {h'_\rho}^1
 +b_\pi g_\pi h_\pi.
 \label{AD-s}
 \end{eqnarray}
 In the ideal case, having the asymmetries at six energy points
 and using the energy dependence
 of $a_i$ and $b_i$ one extract $h_i$ unambiguously.
 In practice, the number of "independent" equations
 for determination of $h_i$ is smaller, because some of
 $a_i$ ($b_i$) are rather weak.
 The energy dependence of the coefficients $a_i$ and $b_i$ is
 shown in the Figs.~\ref{fig:10}(a) and (b), respectively.
 For simplicity, we display only the dominant terms.

 There are several points,
 where $A_{RL}$ and $A_{D}$ are particularly interesting.
 At  $\Delta E\to 0$, where the absolute values
 of  both the asymmetries  have a maximum,
 we get the following relations
\begin{eqnarray}
 A_{RL}&\simeq & -( 4.82 g_\rho h_\rho^0
 + 7.43 g_\rho h_\rho^2
 -0.99g_\omega h_\omega^0)\times 10^{-3},
 \label{ALR-s1}\\
 A_{D}&\simeq & - A_{RL}.
 \label{AD-s1}
 \end{eqnarray}.
  The point  $\Delta E\sim10$ MeV can be used for analyzing  the
  $\pi$-meson exchange contribution
  in $A_D$:
\begin{eqnarray}
 A_{D}\simeq
 (1.46 g_\rho h_\rho^0
  - 0.36 g_\rho h_\rho^2
 +0.27 g_\omega h_\omega^0
 -0.43 g_\pi h_\pi)\times 10^{-3}.
 \label{AD-s2}
 \end{eqnarray}
 The coefficient $b_\pi$ is governed by the long range
 interactions and therefore is not sensitive to the model of
 NN-interaction at short distances.

 The position of intercept $A_{RL}=0$ at $\Delta E_\gamma\simeq 1.3$ MeV
 may be also used for fixing
 the relation between coupling constants, but the
 experiment to find this position would be very difficult.
 On the other hand, another relations
 may be obtained when one of the term
 in Eqs.~(\ref{ALR-s}) and (\ref{AD-s}) vanishes but asymmetries have
 a finite and  reasonable value.
 Thus, we have at $\Delta E_\gamma\simeq 0.4$,  $a_\rho^0=0$, and
 therefore
\begin{eqnarray}
 A_{RL}(\Delta E_\gamma\simeq 0.4\,\, {\rm MeV})
 &\simeq & -( 3.13 g_\rho h_\rho^2
 - 0.67 g_\omega h_\omega^0 )\times 10^{-3}.
 \label{ALR-s3}
 \end{eqnarray}

 Relations {(\ref{ALR-s1}) - (\ref{ALR-s3}) are derived
using the energy dependence of the coefficients $a_i$ and $b_i$ in
Eqs.~(\ref{ALR-s}) and~(\ref{AD-s}) shown in Fig.~\ref{fig:10}.
The later is defined by the short range behaviour of NN-forces,
and is obtained with the Paris potential which has been, in
particularly, designed for the adequate description various
phenomena sensitive to the nucleon interaction at short distances.
On the other hand, the Paris potential can not describe the
neutron-proton scattering length which is its obvious
disadvantage.
 {Nevertheless, we convince that our results
for the Paris potential would be coincide within $\sim20-30$\%
accuracy with the predictions obtained with the other soft-core
realistic potentials. This level of accuracy corresponds to the
difference between our result and previous calculations of
$P_\gamma$=$A_{RL}(E_\gamma=E_{\rm thr})$ with different realistic
potentials~\cite{Ptheor_old}.

\section{summary}

 We have analyzed the energy dependence of two PNC-asymmetries in
 the deuteron photo-disintegration: one with circularly polarized
 photon beam  ($A_{RL}$) and another with polarized
 deuteron target ($A_D$). We show that by combining
 the measurements  of $A_{RL}$and $A_{D}$, valuable information
 on the PNC-nuclear forces may be obtained.
 Namely, using the energy
 dependence of $A_{RL}$ and $A_{D}$, three constraints (equations)
 for determination of the PNC-coupling constants.

 Finally, we stress that the present investigation is a very first
 step. It would be important to verify if the
 predicted asymmetries are universal in the framework of other realistic
 potentials invoking the meson-exchange
 currents and relativistic effects~\cite{Karmanov}. The role of
 the higher multipole transitions at higher energy is not quite clear.

  After completing this paper, the work by Liu, Hyun, and Desplanques
  has appeared in arXiv~\cite{LHD}.
 The authors  have analyzed
 the $A_{RL}$-asymmetry using the realistic Argonne AV${18}$ - potential.
 In spite of some difference in our models, the results of both
 papers are consistent to each other. Ref.~\cite{LHD} gives
 $A_{RL}(\Delta E_\gamma\simeq 0)\simeq +2.53\times10^{-8}$,
 and $A_{RL}$ changes its sign at
 $\Delta E_\gamma\sim1.5$ MeV. The contribution of the
 week $\pi$-exchange transition is suppressed dynamically
 and it is about a factor of 30 smaller than the prediction of
 Ref.~\cite{Oka}.

\acknowledgments

 We thank S.~Date', H.~Ejiri, C.-P.~Liu, I.~Khriplovich, and Y.~Ohashi
 for fruitful discussion. One of author (A.I.T.) thanks
 M.~Yasuoko, the director of Advanced Science Research Center,
 for his hospitality to stay at SPring-8.
 This work was supported in part
 by the Japan  Society for the Promotion of Science (JSPS),
 and was strongly stimulated by a new project to produce a
 high-intensity MeV $\gamma$-rays by inverse Compton scattering at
 SPring-8.

 \begin{table}
  \caption{Weak coupling constants determined from the "best
  value" of Ref.~\cite{DDH}. All values are given in units of
  $10^{-6}$}
  \label{tab:1}
 \begin{ruledtabular}
 \begin{tabular}{lccccccc}
 $h^0_\rho $ & $h^1_\rho $ & ${h^1}'_\rho $ & $h^2_\rho $& $h^0_\omega$
 & $h^1_\omega $ &$h_\pi $&
 \\ \hline
  -1.14 &  -0.02 & -0.07 &-0.95 & -0.19& -0.11 &\,0.46  &
\end{tabular}
\end{ruledtabular}
 \vspace*{0.5cm}
\end{table}

\begin{figure}[ht] \centering
  \includegraphics[width=65mm]{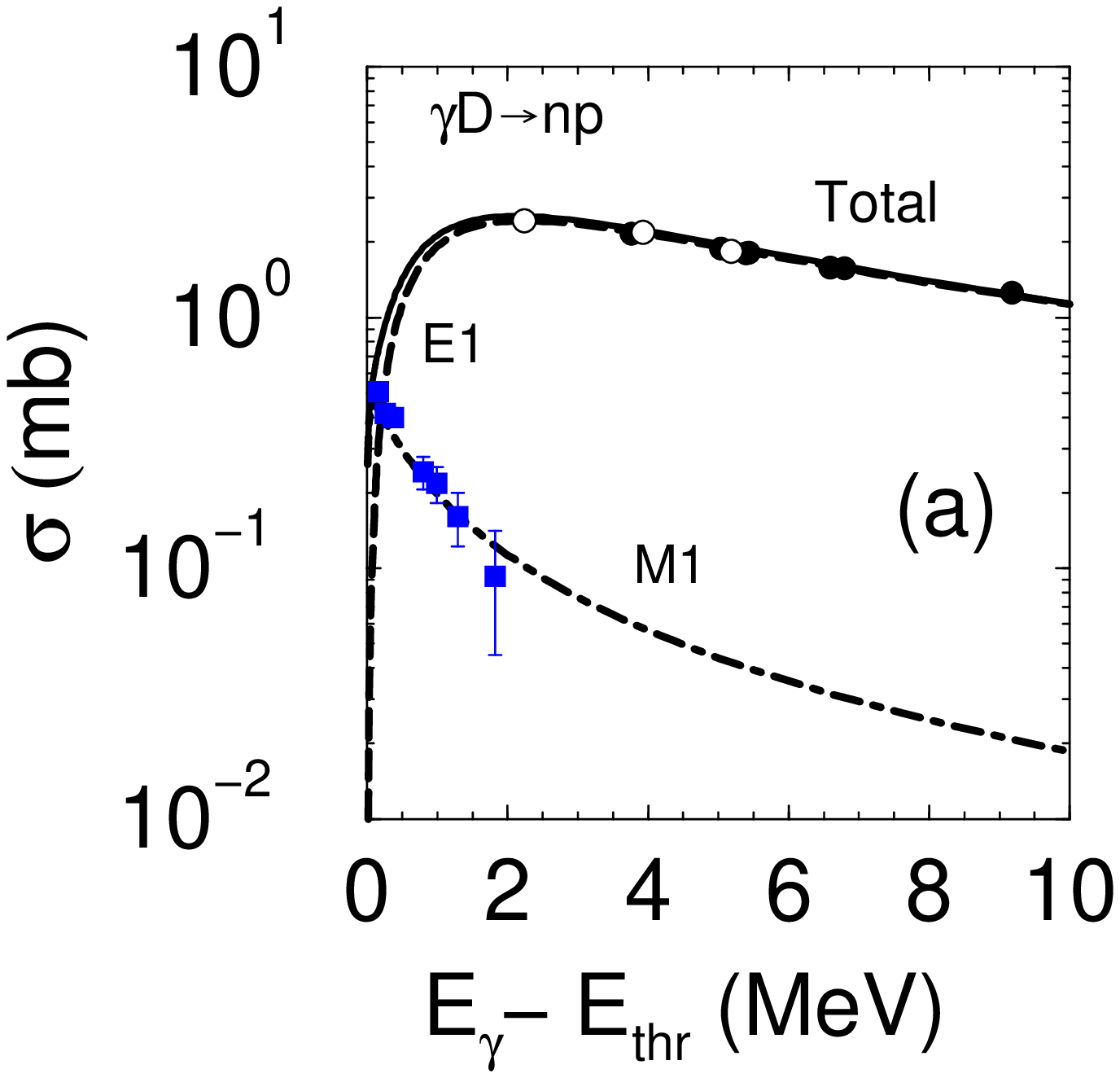}\qquad
 \includegraphics[width=65mm]{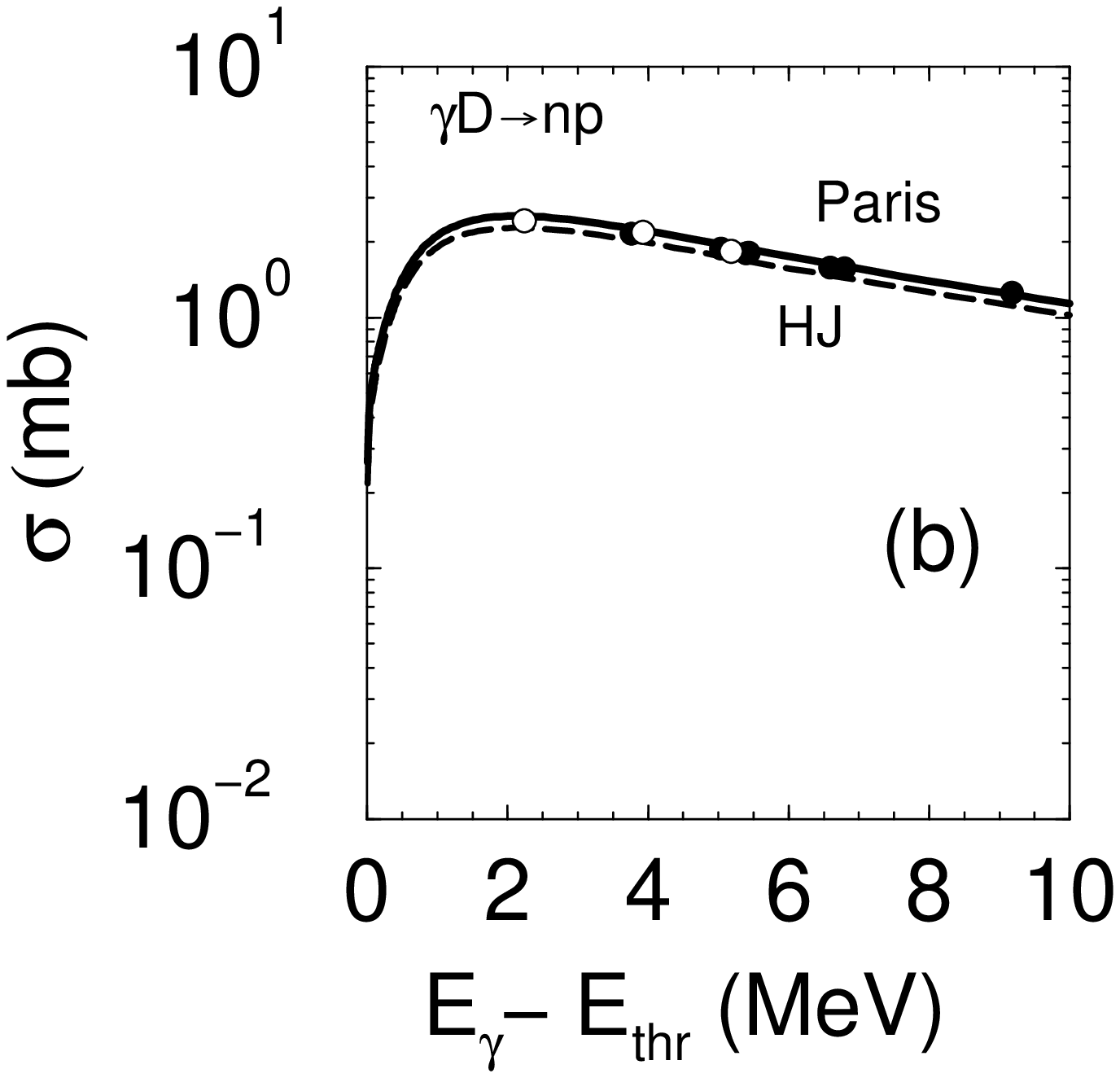}
 \caption{The total cross section of the deuteron
 photo-disintegration as a function of the energy excess
 $\Delta E_\gamma=E_\gamma-E_{\rm thr}$.
 (a) Result for the Paris potential.
 Contributions of the $M1$ and $E1$ transitions are shown by the dashed
 and dot-dashed curves, respectively. (b)~The total cross section for
 the Paris (solid) and Hamada-Johnston (dashed) potentials. The experimental data
 on the total cross section are taken from Refs.~\protect\cite{Barnes} (open circles)
 and~\protect\cite{Birenbaum} (filled circles).
 The data on $M1$-transition (filled squares) are taken
 from Ref.~\protect\cite{Tornow}.}
 \label{fig:2}
\end{figure}
 \begin{figure}[h] \centering
    \includegraphics[width=47.5mm]{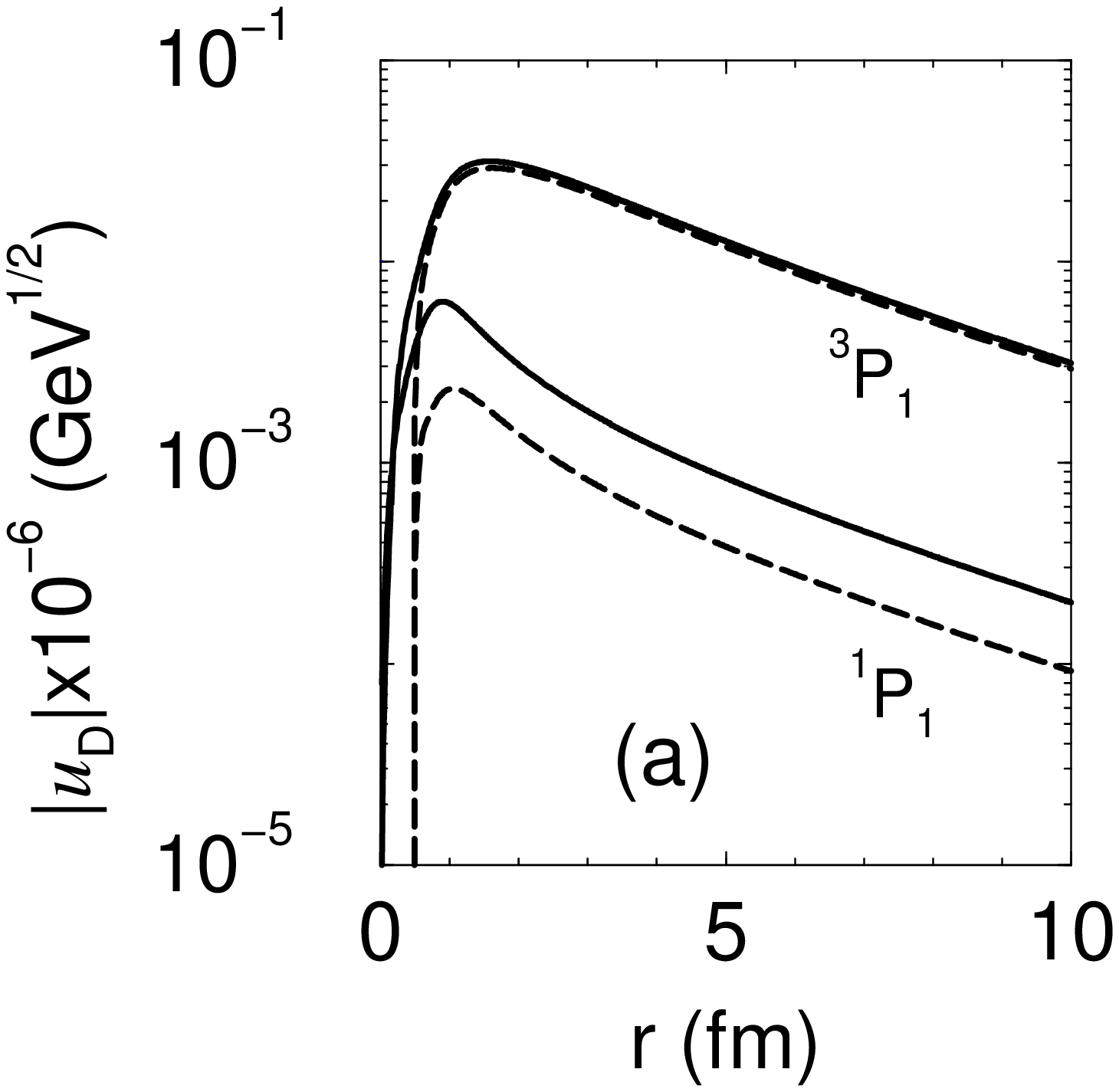}\qquad
    \includegraphics[width=47mm]{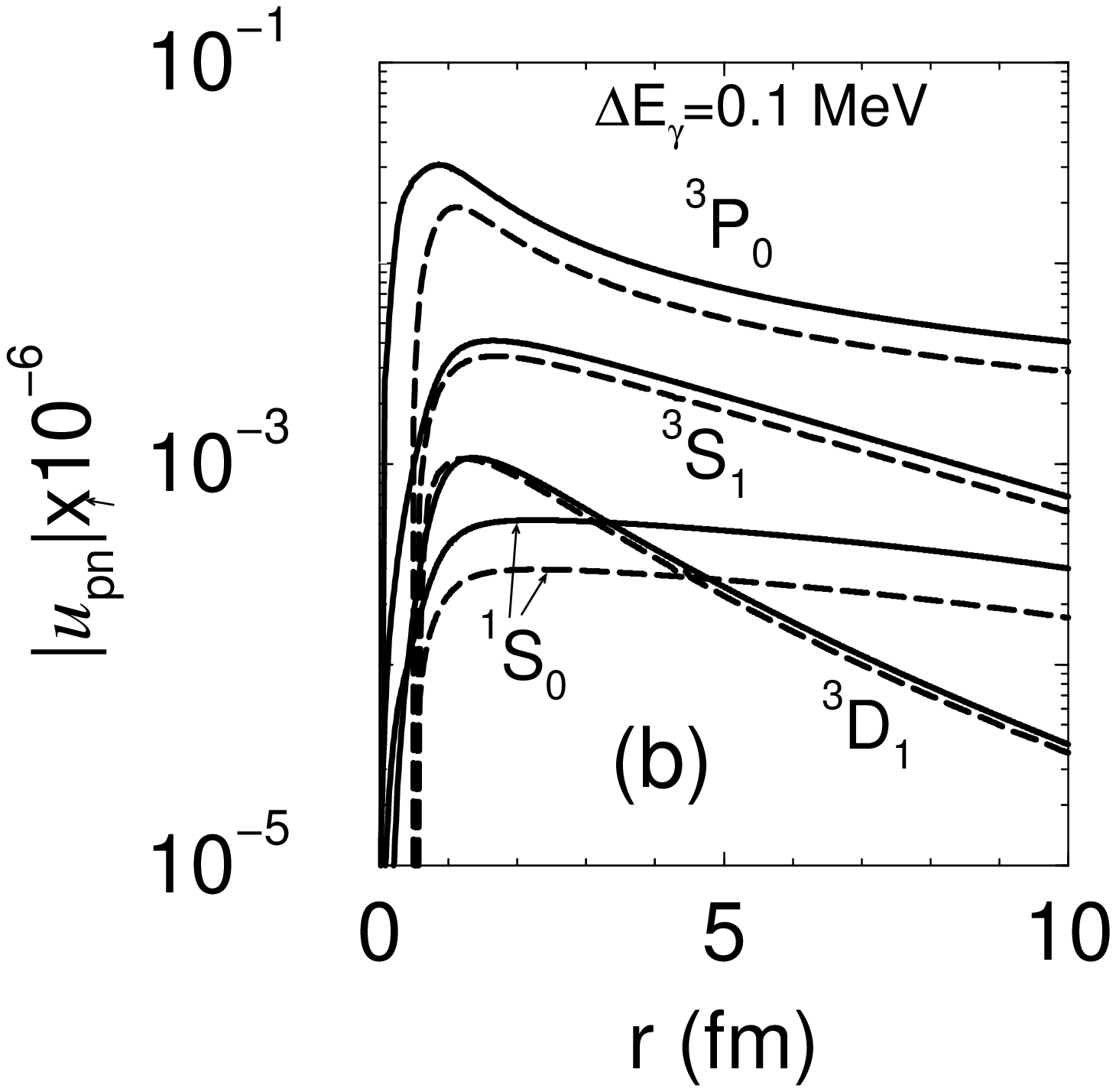}\qquad
   \includegraphics[width=47mm]{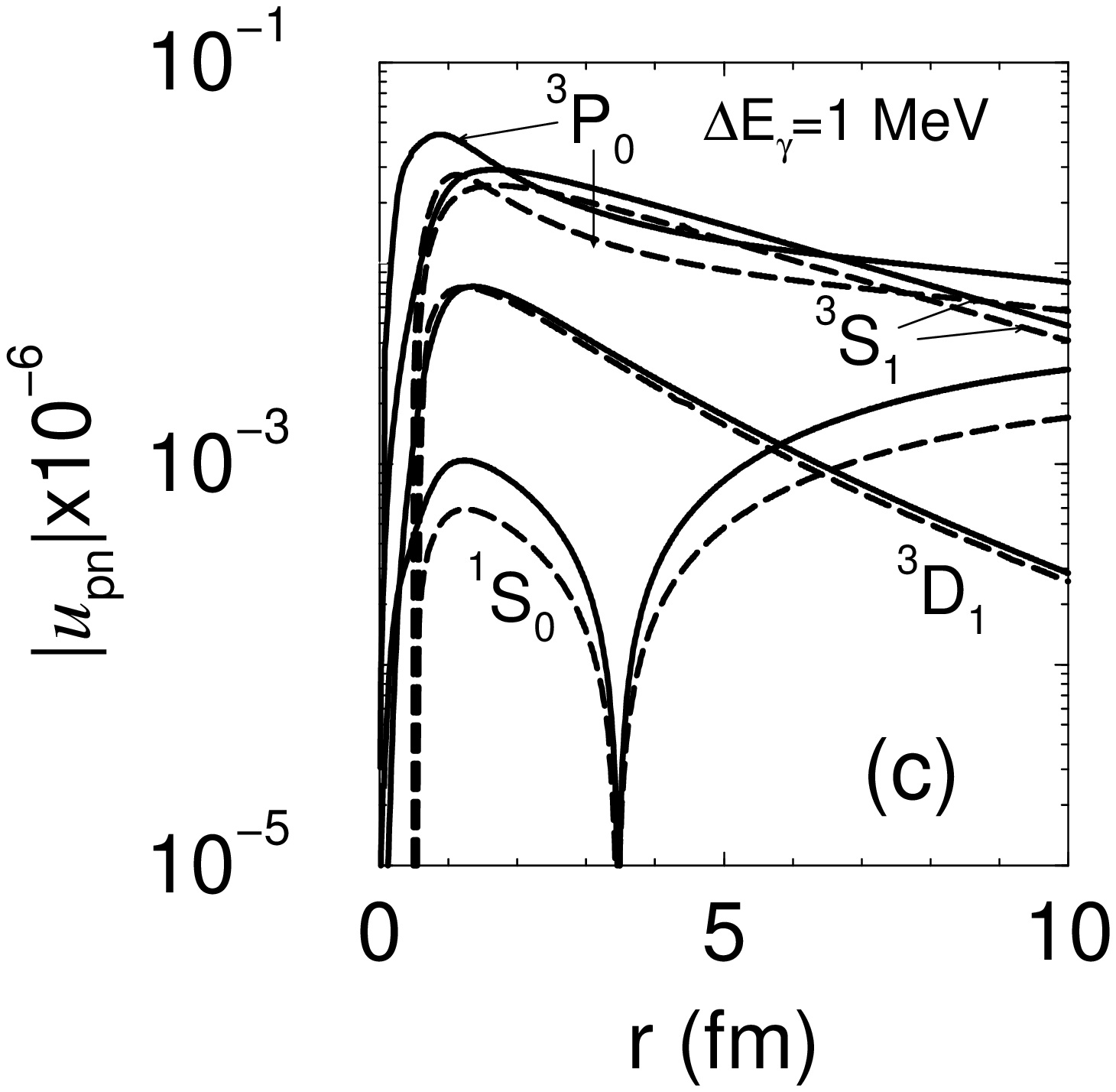}
 \caption{The odd-parity admixture to the proton-neutron wave functions
 calculated with the  Paris (solid curves) and HJ (dashed curves)
 potentials.
 (a) Results for the deuteron wave functions.
 (b) and (c)  Results for the continuum
 $np$- wave functions at $\Delta E_\gamma=0.1$ and 1 MeV, respectively.}
 \label{fig:3}
\end{figure}
\begin{figure}[h] \centering
 \includegraphics[width=65mm]{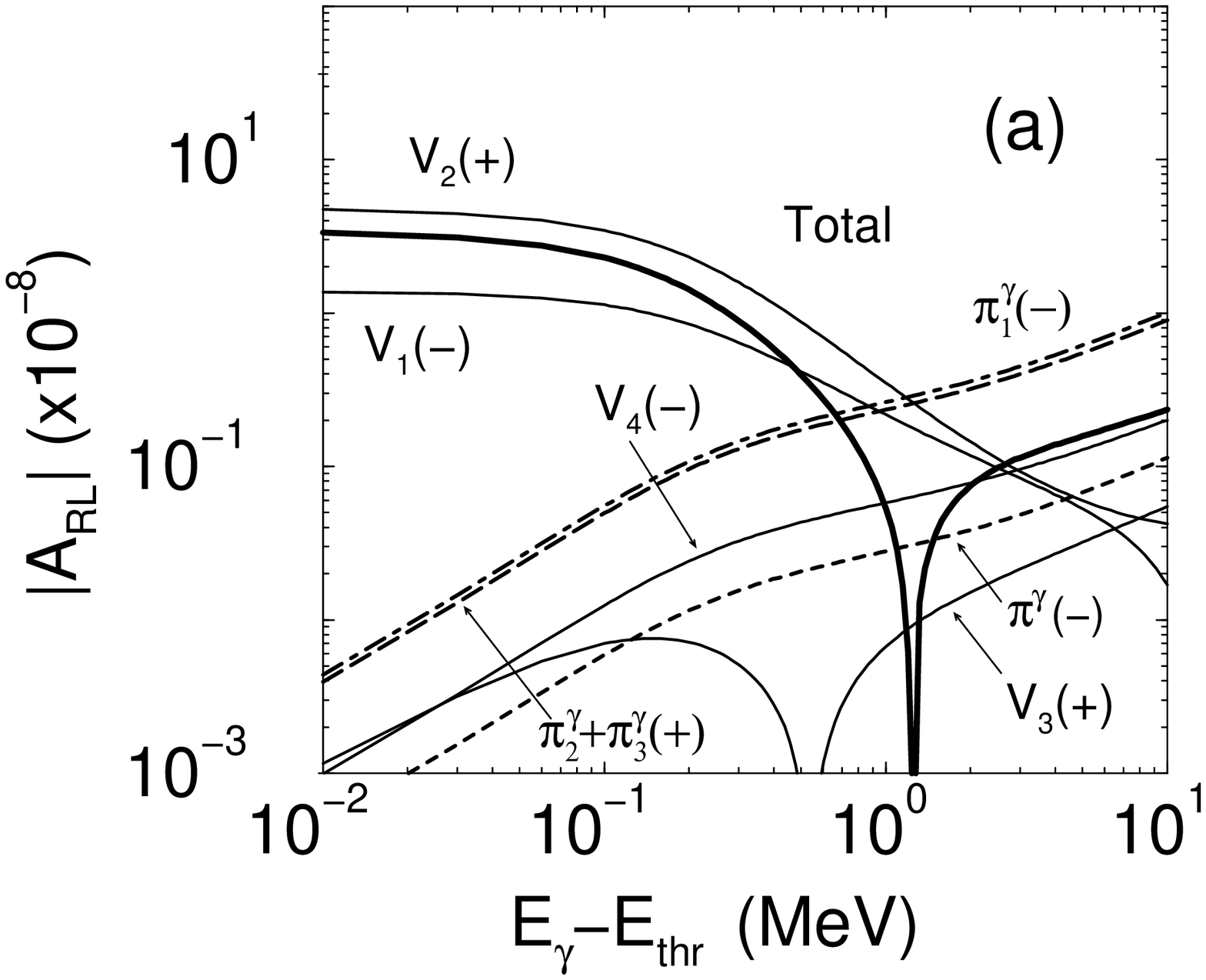}\qquad
 \includegraphics[width=65mm]{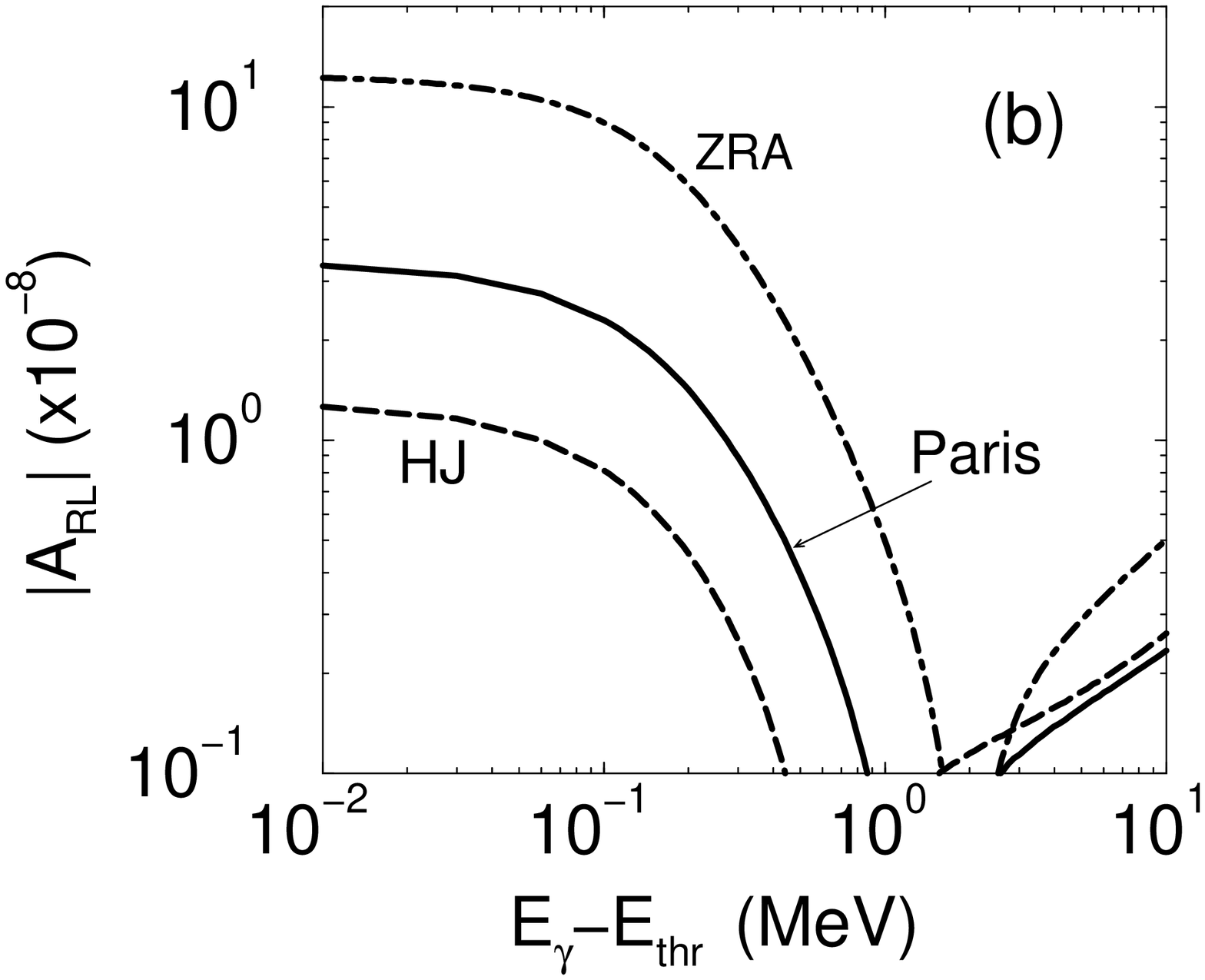}
 \caption{Asymmetry of the deuteron disintegration
 in the reaction $\gamma D\to pn$ with circular polarized
 photon and unpolarized deuteron as a function of
 energy excess $E_\gamma- E_{\rm thr}$.
 (a) Relative  contribution of the  different odd-parity
 transitions for the Paris potential. The sign in the bracket
 denotes the sign of the corresponding term.
 (b) Comparison of the total asymmetry  for the
 Paris (solid), Hamada-Johnston (dashed) potentials and the modified
 ZRA  of Ref.~\protect\cite{KK} (dot-dashed).}
 \label{fig:7}
\end{figure}
\begin{figure}[h] \centering
    \includegraphics[width=65mm]{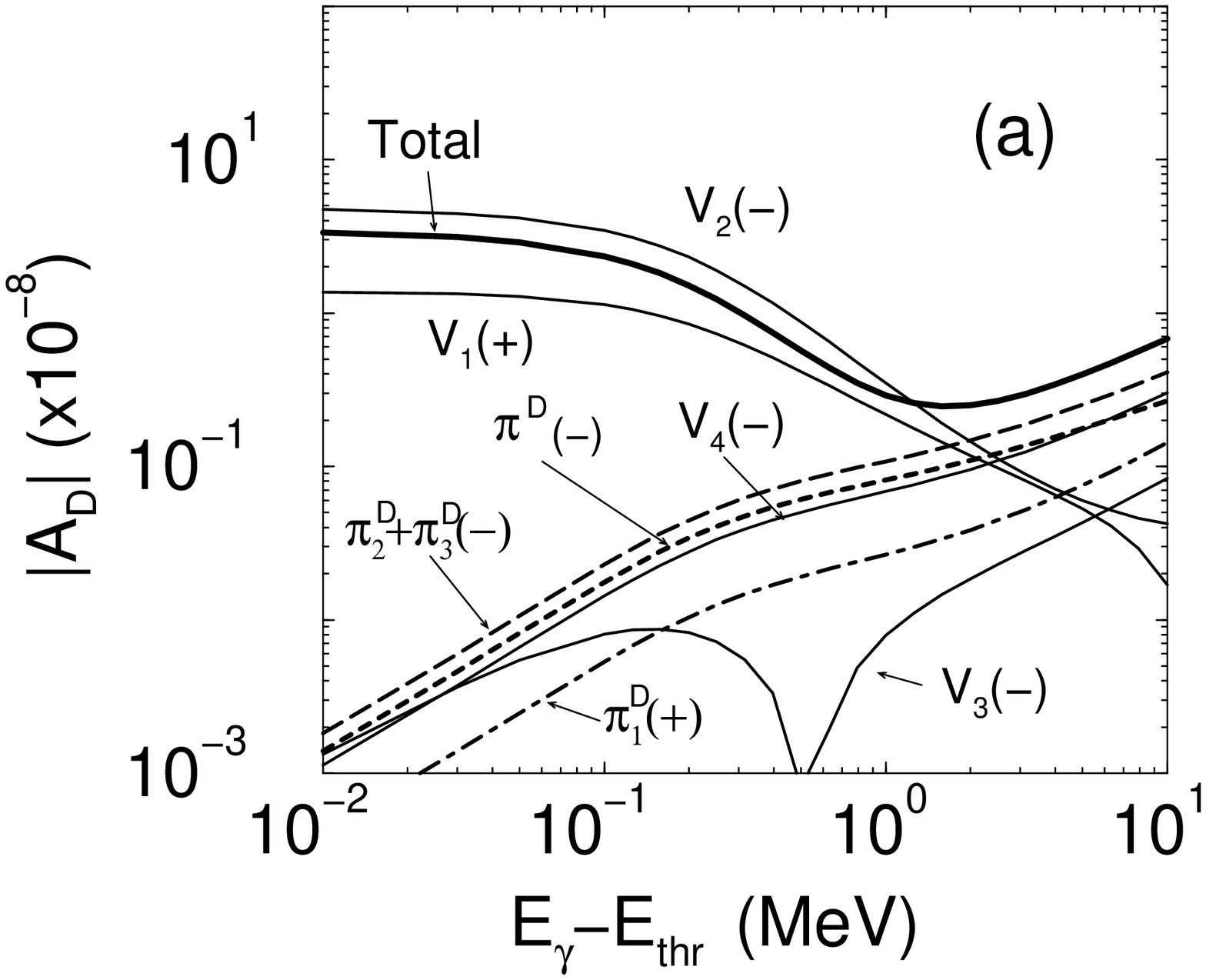}\qquad
    \includegraphics[width=65mm]{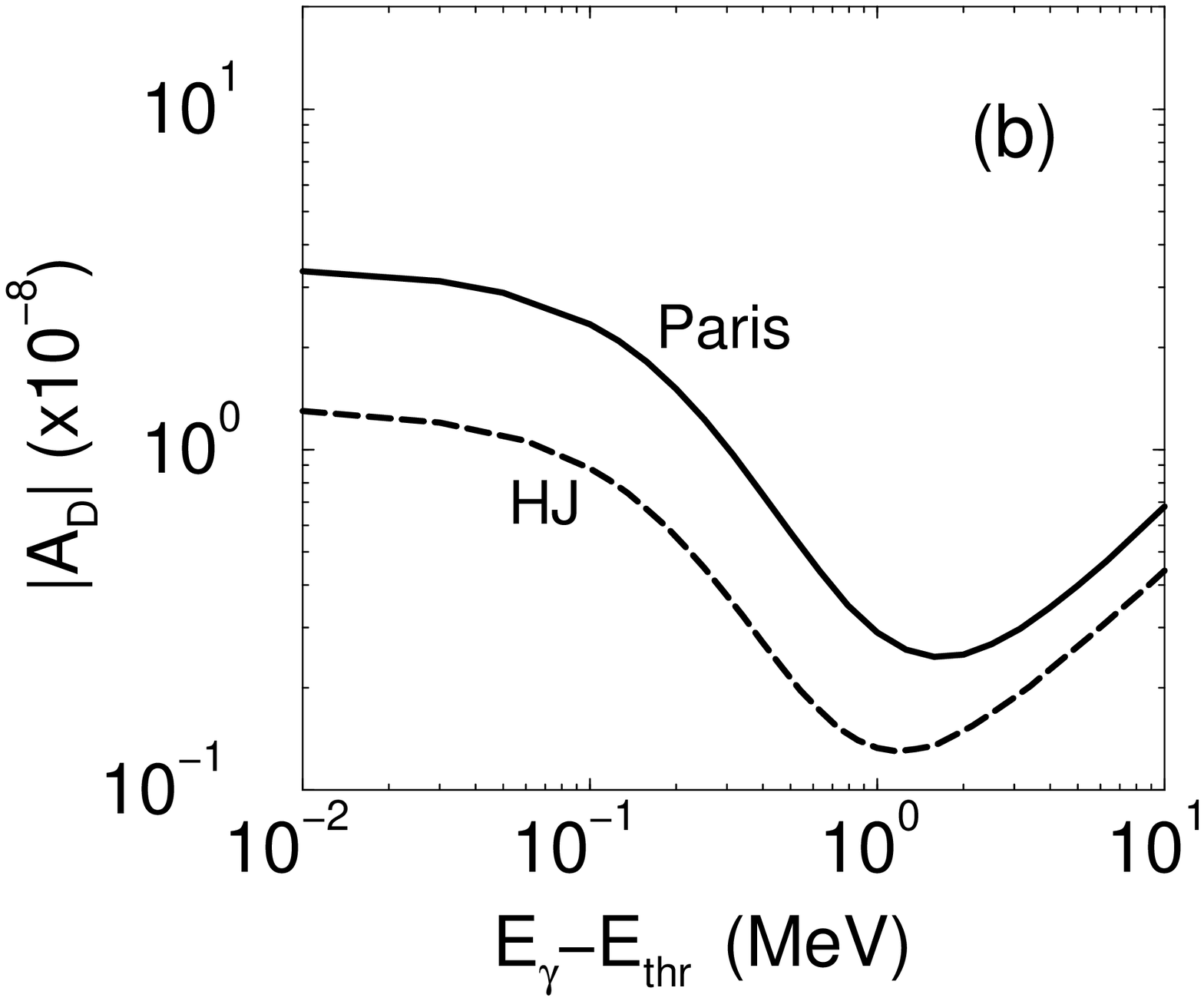}
 \caption{Asymmetry of the deuteron disintegration
 in the  $\gamma D\to pn$ reaction with polarized deuteron and
 unpolarized photon beam as a function of
 energy excess $E_\gamma- E_{\rm thr}$.
 (a) Relative  contribution of different odd-parity
 transitions for the Paris potential. Notation is the same as in
 Fig.~\protect\ref{fig:7}~(a).
 (b) Comparison of the asymmetry  for the
 Paris and  Hamada-Johnston potentials.}
 \label{fig:8}
\end{figure}
\begin{figure}[h] \centering
    \includegraphics[width=65mm]{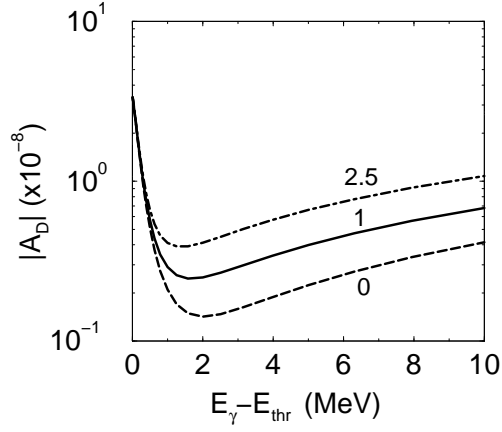}
 \caption{Asymmetry  of the deuteron disintegration
 in the $\gamma D\to pn$  reaction ($A_D$)  with different values of the
 PNC $\pi$-exchange coupling constant:
  $R=f_{\pi}/f_{\pi}^{\rm best}=0,1,2.5$,
  where $f_{\pi}^{\rm best}$ is the "best value" of
  Ref.\protect\cite{DDH}.}
 \label{fig:9}
\end{figure}
\begin{figure}[h] \centering
 \includegraphics[width=65mm]{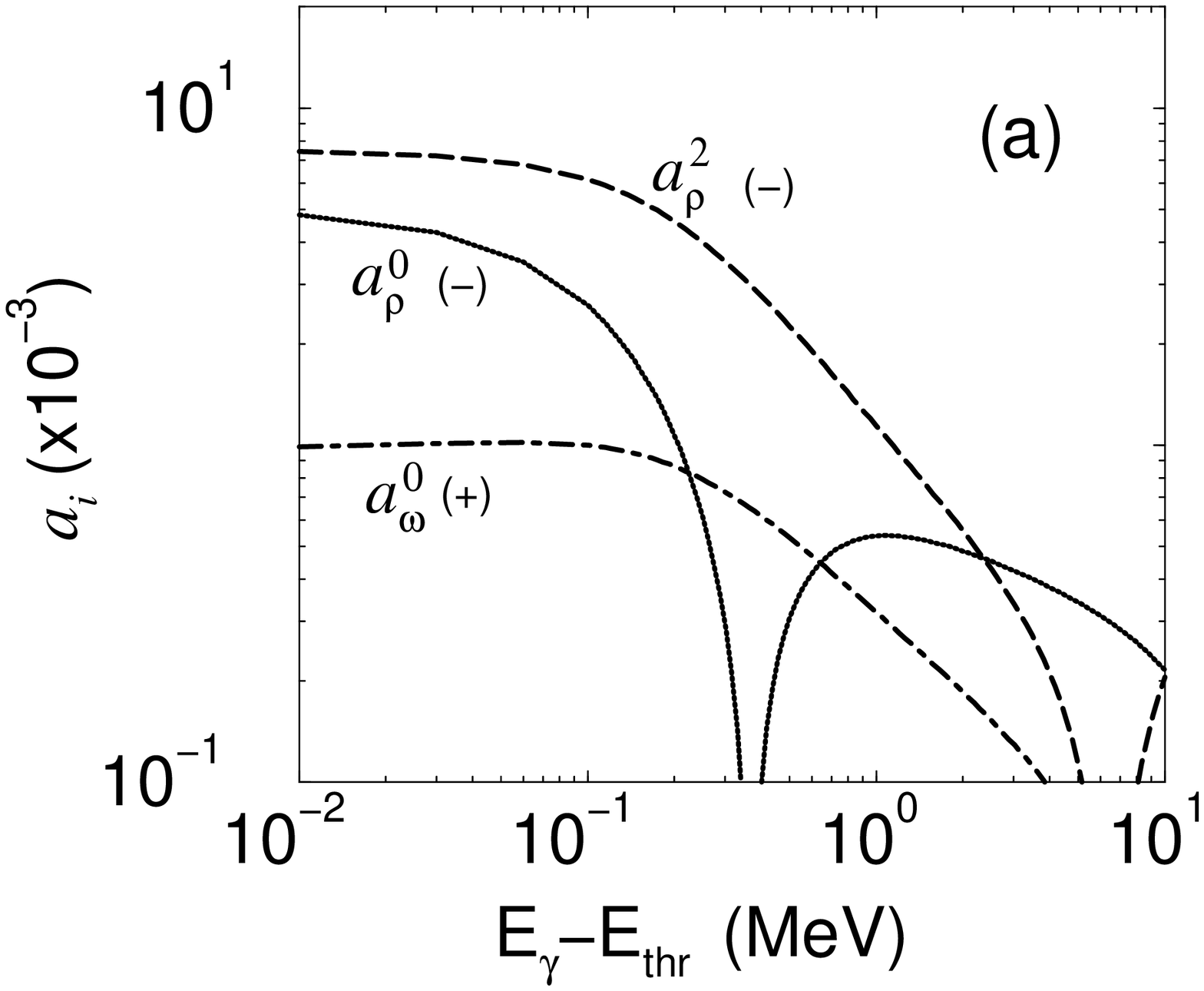}\qquad
 \includegraphics[width=65mm]{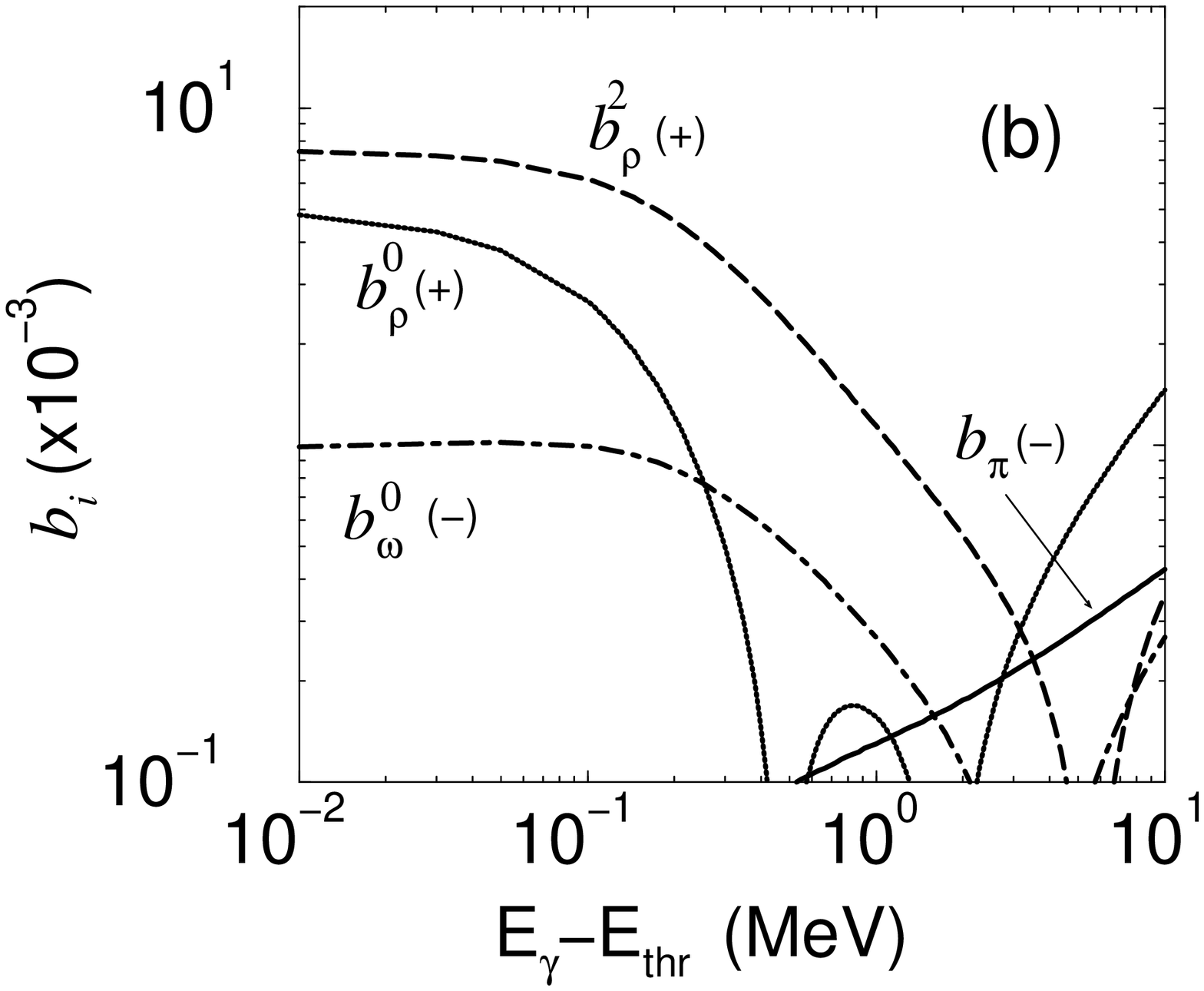}
 \caption{(a) The quantities  $a_i$ of Eq.~(\protect\ref{ALR-s1}).
 (b) The quantities $b_i$ of Eq.~(\protect\ref{AD-s1}).
 We display only the largest components.
 Results are obtained with the Paris potential.}
 \label{fig:10}
\end{figure}

 \end{document}